\documentclass{rspublic}

\usepackage{natbib}
\usepackage{graphicx}
\usepackage{amsmath}
\usepackage{setspace}

\newcommand{\bra}[1]{\langle #1|}
\newcommand{\ket}[1]{| #1 \rangle}
\newcommand{\op}[1]{\hat{#1}}

\newcommand{\cre}[1]{\hat{#1}^\dagger}
\newcommand{\des}[1]{\hat{#1}}
\newcommand{\expi}[1]{ {\rm e}^{{\rm i} #1}}

\begin{document}

\title[Review. Quantum Vortex Nucleation.]{Strong correlations in quantum vortex nucleation of ultracold atomic gases}

\author[Nunnenkamp, Rey and Burnett]{Andreas Nunnenkamp$^{1}$, Ana Maria Rey$^{2}$ and Keith Burnett$^{3}$}
\affiliation{
$^{1}$Departments of Physics and Applied Physics, Yale University, New Haven, CT 06520, USA,
$^{2}$JILA, NIST and Department of Physics, University of Colorado, Boulder, CO 80309, USA,
$^{3}$University of Sheffield, Firth Court, Western Bank, Sheffield, S10 2TN, UK}

\label{firstpage}

\maketitle

\begin{abstract}{Bose-Einstein condensates, optical lattices, vortices, entanglement}
We review some recent developments in the theory of rotating atomic gases. These studies have thrown light on the process of nucleation of vortices in regimes where mean-field methods are inadequate. In our review we shall describe and compare quantum vortex nucleation of a dilute ultracold bosonic gas trapped in three different configurations: a one-dimensional ring lattice, a one-dimensional ring superlattice and a two-dimensional asymmetric harmonic trap. In all of them there is a critical rotation frequency, at which the particles in the ground state exhibit strong quantum correlations. However, the entanglement properties vary significantly from case to case. We explain these differences by characterizing the intermediate states that participate in the vortex nucleation process. Finally, we show that noise correlations are sensitive to these differences. These new studies have, therefore, shown how novel quantum states may be produced and probed in future experiments with rotating neutral atom systems.
\end{abstract}

\section{Introduction}

Vortex nucleation is a topic at the heart of the nature of superfluids and their intrinsic quantum character. Superfluid flow is a direct manifestation of quantum mechanics at the macroscopic level and is only stable below a critical velocity. Above the critical velocity the generation of phonons, rotons (in ${}^4$He) or vortices can lead to a breakdown of superfluidity \citep{Ihas1992}. The microscopic nature of these processes cannot, however, be studied in superfluid liquid helium. There has been a resurgence in the study of vortices  and their production in the new class of superfluids produced using ultracold atoms. Indeed ultracold atoms offer a unique opportunity to investigate topological excitations as recent experiments have demonstrated both with bosonic and fermionic atoms \citep{Madison2000, Raman2001, Haljan2001, Zwierlein2005}. The production of vortex arrays has, in fact, been crucial in demonstrating presence of a superfluid order parameter in the ultracold atom systems. 

Aspects of vortex nucleation in ultracold atoms have been analyzed in various theoretical studies based on a mean-field treatment \citep{Feder1999, Sinha2001, Kasamatsu2003, Lobo2004}. More recently, attempts have been made to approach situations where the neutral atom vortices require methods beyond mean-field techniques. This should be the case, for example, in rapidly rotating clouds where the neutral atom vortices are analogous to those studied in the quantum Hall effect. An overview of the experimental field has been given by \cite{Stock2005}, while \cite{Cooper2009} reviews the different regimes theoretically.

In recent studies, the strongly-correlated nature of the ground state around the region for quantum nucleation has been elucidated. The importance of such studies is at least twofold: firstly production of novel quantum states and secondly investigating the role of strong correlation in the nucleation process.
Two different physical scenarios were considered: In one of them the superfluid ultracold atoms were trapped in an optical ring lattice that can be rotated to introduce angular momentum into the system \citep{Hallwood06a, Hallwood06b, Rey07, Nunnenkamp08, Nunnenkamp2008}; in the other the atoms were confined in a rotating asymmetric trap under conditions equivalent to having charged particles in a magnetic field \citep{Dagnino2009a, Dagnino2009b}.
In the case of the rotating ring lattice the authors were most concerned with the production of Schr\"odinger cat states, i.e.~macroscopic superpositions of states with and without a vortex. They were inspired by the similarities of this system with a superconducting quantum interference device (SQUID) which exhibits macroscopic tunneling between states of opposite current flow \citep{Rouse1995}. As discussed in \cite{Leggett2002}, these macroscopic superposition states are important for testing the limits of validity of quantum mechanics and can be used to achieve quantum-limited measurements in precision spectroscopy \citep{Leibfried2004, Cappellaro2005, Lee2005} which is important for ultra-precise gyroscopes. In \cite{Hallwood06a, Hallwood06b}, \cite{Rey07}, \cite{Nunnenkamp08}, and \cite{Nunnenkamp2008} the goal was to find the optimal conditions for cat state production.
In the case of the vortex nucleation in the lowest Landau level (LLL) \cite{Dagnino2009a, Dagnino2009b} demonstrated that the mean-field picture breaks down close to the nucleation point. The nature of the correlated states, while highly entangled, was clearly distinct from those produced in the rotating ring lattice.

In the present work we show how this difference comes about by studying the nature of the intermediate states, i.e.~the routes that connect states with and without vortices. By constructing effective Hamiltonians, we clarify the competing role of interactions and trap or lattice asymmetry in each of these cases and show that the entangled nature of the ground state and the strong correlations involved in the nucleation process can be seen in quantum noise correlations. We believe that rotating atomic gases offer significant new opportunities to study strongly-correlated atomic systems. They may also have applications to areas where such states can be used in quantum information science and precision measurement.

\section{Vortex nucleation in the rotating ring lattice}

In this section, we investigate the effects of rotation on ultracold bosons confined to one-dimensional ring lattices and superlattices. This is an attractive system for study as it can be produced in the laboratory and isolates important issues of the underlying physics of the nucleation process. We find that at commensurate filling there exists a critical rotation frequency, at which the ground state of the weakly-interacting gas is fragmented into a macroscopic superposition of different quasi-mo\-men\-tum states \citep{Hallwood06a, Hallwood06b, Rey07, Nunnenkamp08, Nunnenkamp2008}.
We note that \cite{Watanabe2007} and \cite{Danshita2009} have studied related aspects in a similar system.

\subsection{Hamiltonian}

We consider a system of $N$ ultracold bosons with mass $M$ confined in a 1D ring lattice of $L$ sites with lattice constant $d$. Optical ring lattices can be created with Laguerre-Gaussian (LG) laser beams, as proposed by \cite{Amico05} and realized by \cite{Franke-Arnold2007}. LG beams can be derived from ordinary Gaussian beams, e.g.~by means of computer-generated phase holograms~\citep{Chavez-Cerda2002}. A conceptionally different approach to arbitrary 2D potentials are spatial light modulators which have recently been used for dynamical manipulation of Bose-Einstein condensates (BECs)~\citep{Boyer2006}.
The ring is rotated in its plane (about the $z$-axis) with angular velocity $\Omega$. By subtracting the rotation energy $\op{H}_\textrm{rot} = \int d\mathbf{x} \,\hat{\Phi}^{\dagger} \Omega \hat{L}_z \hat{\Phi}$ we transform to the rotating frame where the potential $V(\mathbf{x})$ is time-independent and the many-body Hamiltonian is given by~\citep{Bhat2006, Rey07}
\begin{equation}
\hat {H}=\int d\mathbf{x}
\hat{\Phi}^{\dagger}\left[-\frac{\hbar^2}{2 M}\nabla^2 +
V(\mathbf{x}) +\frac{4 \pi \hbar^2 a_s}{2M}
\hat{\Phi}^{\dagger}\hat{\Phi}- \Omega \hat{L}_z\right]\hat{\Phi}.
\label{generalHam}
\end{equation}
In this expression, $a_s$ is the $s$-wave scattering length, $V(\mathbf{x})$ the lattice potential, $\hat{L}_z$ the angular momentum and $\mathbf{x}$ the 3D spatial coordinate vector. $\hat{\Phi} (\mathbf{x})^{\dagger}$ and $\hat{\Phi} (\mathbf{x})$ are bosonic creation and annihilation field operators.

We assume that the lattice potential $V(\mathbf{x})$ confines the motion along the $z$-axis as well as the radial direction in the $x-y$ plane so strongly that only the motion of the atoms along the ring has to be taken into account. In addition, we assume that the lattice is deep enough to restrict tunneling to nearest-neighbor sites and that the band gap is larger than the rotational energy. These assumptions imply that the bosonic field operator $\hat{\Phi}$ can be expanded in Wannier orbitals confined to the first band $\hat{\Phi}(\mathbf{x}) = \sum_j \hat{a}_j W_j'(\mathbf{x})$ \citep{Jaksch98}. Here, $W_j'(\mathbf{x})$ are the Wannier orbitals in the rotating frame and $\hat{a}_j$ the bosonic annihilation operator of a particle at site $j$.
We recall that the Hamiltonian of a neutral particle in a frame rotating at frequency $\Omega$ around the $z$-axis, $\op{H} = \mathbf{\op{p}}^2/2M - \Omega L_z$, is equivalent to the Hamiltonian of a charged particle in a magnetic field along the $z$-axis, $\op{H} = (\mathbf{\op{p}} - \mathbf{A})^2/2M$ with the effective vector potential $\mathbf{A}(\mathbf{x}) = M \Omega (\hat{z} \times \mathbf{x})$. This implies that we can first calculate the Wannier orbitals of the stationary lattice $W_j(\mathbf{x})$ and then account for the presence of the effective vector potential $\mathbf{A}(\mathbf{x})$ via the gauge transformation $W_j'(\mathbf{x}) = \exp \left[ \frac{-i}{\hbar} \int_{\mathbf{x}_j}^\mathbf{x} \mathbf{A} (\mathbf{x}') \cdot d\mathbf{x}' \right] W_j(\mathbf{x})$.

In terms of these quantities and up to on-site diagonal terms, the many-body Hamiltonian can be written as~\citep{Bhat2006, Rey07}
\begin{equation}
\op{H}
= -\sum_{j=1}^L \left( J_j \expi{\theta} \cre{a}_{j+1} \des{a}_j + H.c. \right) + \frac{U}{2} \sum_{j=1}^L \hat{n}_j(\hat{n}_j-1).
\label{ham}
\end{equation}
In this expression, $\hat{n}_j=\hat{a}_j^{\dagger}\hat{a}_{j}$ is the number operator at site $j$, $\theta$ is the effective phase twist~\citep{Peierls1933} induced by the gauge field, $\theta \equiv \int_{\mathbf{x}_{i}}^\mathbf{x_{i+1}} \mathbf{A}(\mathbf{x}')\cdot d\mathbf{x}' = \frac{M \Omega L d^2}{h}$, $J_j$ is the hopping energy between nearest-neighbor sites $j$ and $j+1$: $J_j \equiv \int d\mathbf{x} W_j^{*} \left[-\frac{\hbar^2}{2M} \nabla^2 + V(\mathbf{x})\right] W_{j+1}$, and $U$ the on-site interaction energy: $U\equiv \frac{4 \pi a_s \hbar^2}{M} \int d\mathbf{x} \, |W_j|^4$.

\subsection{Superposition states in rotating ring lattices}

We start our discussion with the case of a uniform ring lattice, i.e.~$J_j = J$ for all $j$. To understand the effect of rotation on the atoms in the ring lattice, we write the many-body Hamiltonian (\ref{ham}) in terms of the quasi-momentum operators $\hat{b}_q = \frac{1}{\sqrt{L}} \sum_{j=1}^{L} \hat{a}_j e^{-2\pi i q j/L}$, where $2\pi q / d L$ is the quasi-momentum and $q=0,\dots,L-1$ an integer. In this basis the Hamiltonian (\ref{ham}) has the form~\citep{Hallwood06a, Rey07}
\begin{equation}
\hat{H} = \hat{H}_{\textrm{sp}} + \hat{H}_{\textrm{int}} =
\sum_{q=0}^{L-1} E_q\hat{b}_q^{\dagger}\hat{b}_{q}+\frac{U}{2L}
\sum_{q,s,l=0}^{L-1}\hat{b}_q^{\dagger} \hat{b}_s^{\dagger}
\hat{b}_l\hat{b}_{[q+s-l] \, \textrm{mod} \, L}
\label{momehat}
\end{equation}
where $E_q = -2J \cos (2 \pi q/L - \theta) $ are the single-particle energies, and the modulus is taken because in collision processes the quasi-momentum is conserved up to an integer multiple of the reciprocal lattice vector $2\pi/d$, i.e.~modulo Umklapp processes.

Following \cite{Hallwood06a} and \cite{Rey07}, we show in Fig.~\ref{fig:spspectrum} the single-particle spectrum as a function of the phase twist $\theta$. In the absence of rotation, i.e.~$\theta=0$, the state with zero quasi-momentum $\ket{q=0}$ is the single-particle ground state. In the rotating system the ground state depends on the phase twist $\theta$. Writing $\theta = \frac{2\pi}{L} m + \frac{ \Delta\theta}{L}$ with $m$ an integer and $0\leq \Delta\theta< 2\pi$, the ground state is the quasi-momentum state $\ket{q=m}$ for $ 0\leq \Delta\theta< \pi$ and $\ket{q=m+1}$ for $ \pi< \Delta\theta< 2\pi$.

In the absence of interactions $U = 0$, the ground state of a bosonic many-body system is the state with all $N$ bosons occupying the lowest-energy single-particle state. At rotation frequencies corresponding to phase twists with $\Delta \theta = \pi$, the ground state of the single-particle Hamiltonian $\op{H}_\textrm{sp}$ is two-fold degenerate, i.e.~$E_m = E_{m+1}$, so that there is a $N+1$-dimensional degenerate subspace at the $N$-particle level (see Fig.~\ref{fig:spectrum}). A convenient basis for this subspace are the Fock states $|n,N-n\rangle$ with $0 \le n \le N$, where $n$ particles are in the quasi-momentum state $\ket{q = m}$ with energy $E_m$ and $N-n$ particles in the quasi-momentum state $\ket{q = m+1}$ state with energy $E_{m+1}$, respectively.

\begin{figure}
\centering
\includegraphics[width=\textwidth]{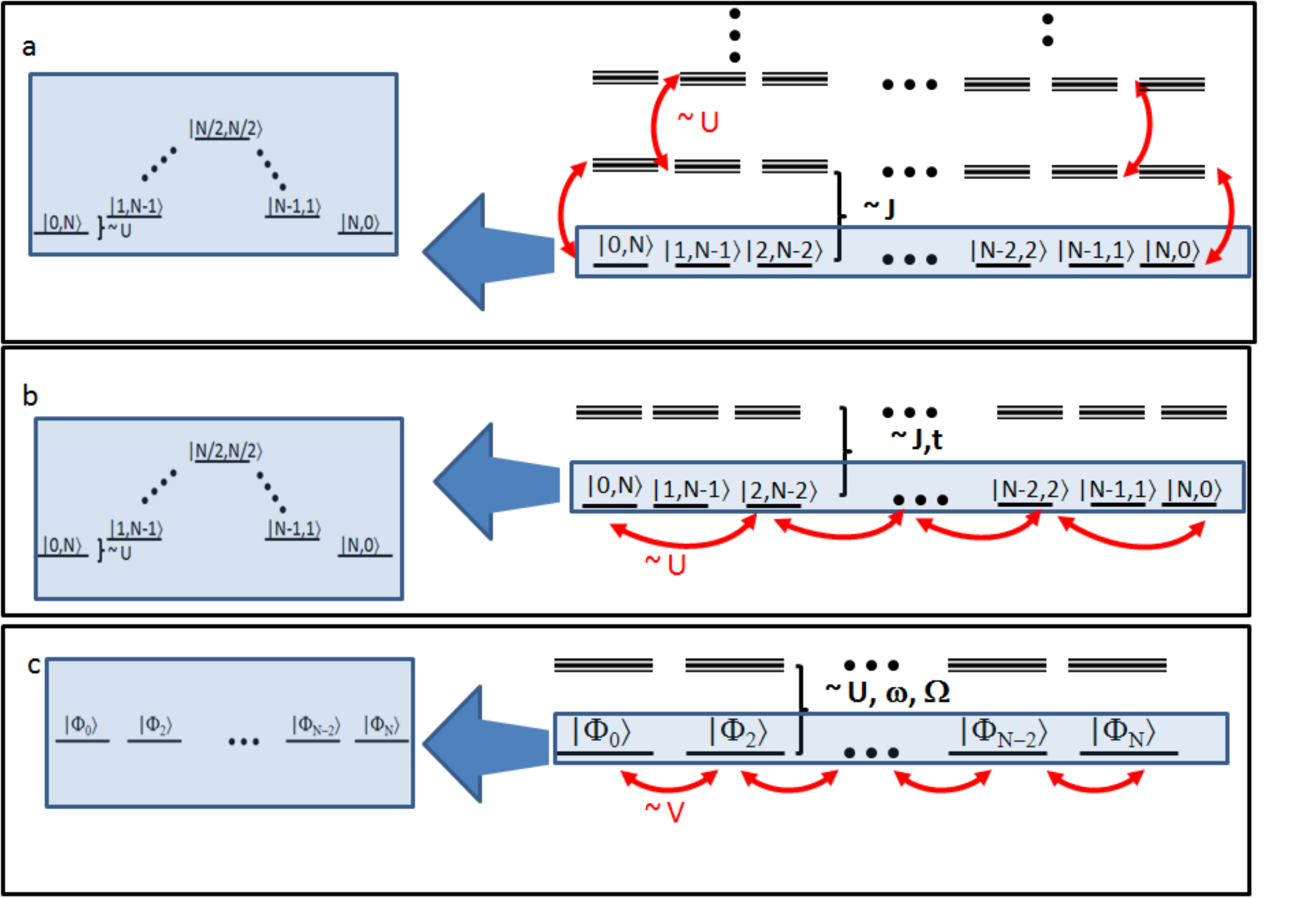}
\caption{{\bfseries Coupling path.} The degeneracies of the many-particle spectrum are lifted by the interactions $U$ in case of the ring lattice and superlattice (upper and middle respectively) or by the asymmetric potential $V$ in case of the LLL system (lower). While in the ring lattice the degenerate states are coupled via non-resonant states \citep{Hallwood06b}, i.e.~states separated by an energy gap $J$, both in the superlattice and LLL cases interactions or asymmetry couple the degenerate states within themselves. In the superlattice case however there are diagonal couplings which introduce small energy gaps within the resonant manifold making the $|N,0\rangle$ and $|0,N\rangle$ the lowest-lying states \citep{Nunnenkamp2008}. These considerations explain why in the lattice cases the ground state is a macroscopic superposition of two quasi-momentum states while in the LLL system it is a more complicated superposition state.}
\label{fig:coupling}
\end{figure}

For weak interactions, i.e.~$U \ll J$, we can use first-order perturbation theory to account for the effect of interactions. It predicts that the energies of the states $|n,N-n\rangle$ are
\begin{equation}
E_{n}^{(1)} = n E_m + (N-n) E_{m+1} + \frac{U}{2L} \left[ N(N-1) + 2n(N-n) \right].
\label{shifted}
\end{equation}
We see from this expression that the degeneracy is lifted and the states of lowest energy are $|N,0\rangle$ and $|0,N\rangle$. These two states are still degenerate and higher-order coupling is needed to break the degeneracy (see Fig.~\ref{fig:coupling}).

At this point it is important whether the number of atoms $N$ is commensurate or incommensurate with the number of lattice sites $L$. While there are many different paths that couple the states $|N,0\rangle$ and $|0,N\rangle$ in the commensurate case, in the incommensurate case there is no coupling between these two states and they thus remain degenerate at all orders of perturbation theory.
To understand this fact, let us consider the total quasi-momentum operator $\hat{K} = \frac{2\pi}{L} \left| \sum _{q=0}^{L-1} q \hat{b}_q^\dagger \hat{b}_q \right|_{\, \textrm{mod} \, L}$. Since the many-body Hamiltonian (\ref{momehat}) commutes with the quasi-momentum operator $\op{K}$, i.e.~$\left[ \op{H},\op{K} \right] = 0$, the Hamiltonian has block diagonal form if the quasi-momentum Fock states are ordered according to the eigenvalues of $\op{K}$. In the commensurate case, we have $N= \bar{n} L$ with the number density $\bar{n}$ being an integer, so that the states $|N,0\rangle$ and $|0,N\rangle$ have total quasi-momentum $K=\frac{2\pi}{L} |m \bar{n} L |_{\, \textrm{mod} \, L} = 0$ and $K=\frac{2\pi}{L} |(m+1) \bar{n} L|_{\, \textrm{mod} \, L} = 0$ with $m$ integer, respectively. Thus both of them belong to the $K=0$ block and are coupled by the interactions. On the other hand, in the incommensurate case, we have $N = \bar{n} L + \Delta N$, and hence the two states $|N,0\rangle$ and $|0,N\rangle$ belong to different blocks and remain degenerate as $K = |m \bar{n} L+ m\Delta N|_{\, \textrm{mod} \, L} \neq |(m+1) \bar{n} L + (m+1) \Delta N |_{\, \textrm{mod} \, L}$.

In the commensurate case, we can construct an effective $2 \times 2$ Hamiltonian by projecting the many-body Hamiltonian (\ref{momehat}) onto the subspace spanned by the states $|N,0\rangle$ and $|0,N\rangle$
\begin{equation}
\op{H}_{2 \times 2} =
\left( \begin{array}{cc}
E_{0}^{(1)}& \Delta \\
\Delta & E_{N}^{(1)} \\
\end{array} \right)
\end{equation}
where the coupling $\Delta$ is given in perturbation theory by
\begin{equation}
 \Delta = \sum_{i,j,\dots p} \frac{H_{0i} H_{ij} \dots H_{pN}}{(E_{0}^{(1)} - \varepsilon_i) (E_{0}^{(1)}-\varepsilon_j) \dots ( E_0^{(1)} - \varepsilon_p)} \propto \left( \frac{U}{2L} \right)^{\bar{n}(L-1)} \label{gap}.
\end{equation}
In this expression, $\bar{n} = N/L$ is the number density, the $H_{ij}$ are transition matrix elements introduced by the interaction Hamiltonian $\op{H}_{\textrm{int}}$, and $\varepsilon_i$ are either given by $E^{(1)}_n$ in Eq.~(\ref{shifted}) or non-interacting many-body eigenenergies depending upon whether the intermediate states are or are not in the $|n,N-n\rangle$ manifold. The factor ${\bar{n}(L-1)}$ corresponds to the minimum number of collision processes necessary to couple the states $|N,0\rangle$ and $|0,N\rangle$ and the sum is taken over all possible coupling paths.
In the case of the ring lattice the coupling is exclusively though states outside the degenerate manifold (see Fig.~\ref{fig:coupling}), and the coupling $\Delta$ decreases exponentially with increasing number of particles $N$.

At the critical phase twist $\Delta \theta = \pi$, we have $E_{0}^{(1)} = E_{N}^{(1)}$ and due to the non-zero value of the coupling $\Delta$ the symmetric and anti-symmetric superpositions $| \pm \rangle$ become the ground and first excited state separated by an energy gap $2 \Delta$
\begin{equation}
| \pm \rangle = \frac{|N,0\rangle \pm |0,N\rangle}{\sqrt{2}}.
\label{cat}
\end{equation}
In Fig.~\ref{fig:overlap} we plot the overlap with the states $\ket{N,0}$ and $\ket{0,N}$. Below and above the critical phase twist the bosons form the non-rotating condensate $\ket{N,0}$ and the vortex state $\ket{0,N}$, respectively, while at resonance the macroscopic superposition state (\ref{cat}) occurs.
These Schr\"odinger cat states are central to high-precision spectroscopy, amplified quantum detection and measurement \citep{Leibfried2004, Cappellaro2005, Lee2005} where they improve the resolution by a factor of $\sqrt{N}$ with respect to the classical shot noise limit.

\subsection{Superposition states in rotating ring superlattices}

In \cite{Nunnenkamp08} we further explored whether the situation can be improved by introducing a lattice modulation and considered instead a ring superlattices where $J_j = J$ for $j$ even and $J_j = t$ for $j$ odd.
Due to the superlattice potential the quasi-momentum states $\ket{q}$ and $\ket{q+L/2}$ are coupled and the single-particle Hamiltonian is no longer diagonal in the quasi-momentum basis
\begin{equation}
\op{H}_{\textrm{sp}}
= \sum_{q=0}^{L/2-1}
\left( \begin{array}{cc} \cre{b}_q & \cre{b}_{q+L/2} \end{array}\right)
\left( \begin{array}{cc}
- (J+t) \cos \phi & - i (J-t) \sin \phi \\
+ i (J-t) \sin \phi & + (J+t) \cos \phi
\end{array} \right)
\left( \begin{array}{c} \des{b}_q \\ \des{b}_{q+L/2} \end{array}\right)
\label{hamsp}
\end{equation} with $\phi=\theta-\frac{2\pi q}{L}$.
We can diagonalize the single-particle Hamiltonian (\ref{hamsp}) via a unitary basis transformation $( \op{c}_q , \op{c}_{q+L/2} ) = {\bf{M_U}} (\op{b}_q , \op{b}_{q+L/2} )$
and obtain
\begin{equation}
\op{H}_{\textrm{sp}}
= \sum_{q=0}^{L/2-1}
\left( \begin{array}{cc} \cre{c}_q & \cre{c}_{q+L/2} \end{array}\right)
\left( \begin{array}{cc}
E_{q}^- & 0 \\
0 & E_{q}^+
\end{array} \right)
\left( \begin{array}{c} \des{c}_q \\ \des{c}_{q+L/2} \end{array}\right)
\label{hamspdiag}
\end{equation}
where the single-particle energies are given by
$E_{q}^{\pm} = \pm \sqrt{J^2 + t^2 + 2J t \cos \left(2\theta-\frac{4\pi q}{L}\right)}$.

In the uniform ring, $t/J=1$, the eigenstates of the single-particle Hamiltonian $\op{H}_{\textrm{sp}}$ are quasi-momentum states. At certain phase twists $\theta$ they are doubly degenerate. For example, for $L=4$ sites the quasi-momentum states $\ket{q=1}$ and $\ket{q=-1}$ are degenerate at $\theta = 0$, whereas at $\theta=\pi/4$ the states $\ket{q=0}$ and $\ket{q=1}$ as well as $\ket{q=2}$ and $\ket{q=-1}$ are degenerate. Reducing the symmetry of the ring by choosing $t\not=J$ the quasi-momentum states which differ by $L/2$ quasi-momentum units are coupled by the single-particle Hamiltonian (\ref{hamsp}), so that the quasi-momentum states $\ket{q=1}$ and $\ket{q=-1}$ hybridize and the degeneracy at $\theta = 0$ is lifted, i.e.~$E_1^+ - E_1^- = 2(J-t)$.

At $\theta = \pi/4$ however the degenerate quasi-momentum states are not coupled by the single-particle Hamiltonian (\ref{hamsp}), so that the degeneracy is present also in the non-uniform case. This remains true for arbitrary $L$, i.e.~$E_0^- = E_{L/4}^-$ at $\theta = \pi/4$, but for $L \not= 4$ these states are not the ground states of the system. In Fig.~\ref{fig:spspectrum} we plot the single-particle spectrum for $L=4$ sites as a function of the effective phase twist $\theta$. It shows level crossings at $\theta=\pi/4$ both for $t = J$ as well as $t\neq J$. We will refer to $\theta=\pi/4$ as the critical phase twist, since -- as we will demonstrate below -- weak on-site interactions lift the degeneracy at $\theta=\pi/4$ and lead to the formation of strongly-correlated states in the many-body system.

The crossing of two single-particle levels implies a $(N+1)$-fold degeneracy in the non-interacting many-body spectrum. In Fig.~\ref{fig:spectrum} we plot the many-body spectrum with $L=N=4$ and $t/J=0.7$ for $U/J=0$ and $U/J=0.5$ as a function of the phase twist $\theta$ and find that interactions turn level crossings into avoided crossings.

In the weakly interacting regime this effect can be understood by deriving an effective Hamiltonian within the $(N+1)$-dimensional degenerate subspace. A convenient basis for this subspace is spanned by the Fock states $|n,N-n\rangle$ with $0 \le n \le N$, where $n$ particles are in the single-particle state of energy $E_0^{-}$ and $N-n$ particles in the one of energy $E_{L/4}^{-}$, respectively. For weak interactions $NU/L \ll 2\sqrt{J^2+t^2}$ this subspace is the low-energy sector of the many-body problem for all phase twists $\theta$ and tunneling strength ratios $t/J$.
Starting from the interaction Hamiltonian (\ref{momehat}) we restrict the Hilbert space to the relevant modes and keep only terms within the low-energy subspace. In this way we obtain the effective Hamiltonian to first order in the on-site interaction strength $U$
\begin{equation}
\op{H}_{\textrm{eff}} = \left(E_{0}^- \op{n}_0 + E_{L/4}^- \op{n}_{L/4} \right)
+ \frac{U}{2L} \left( 2 \op{n}_0 \op{n}_{L/4} + N^2 - N \right)
+ \left( \frac{i \eta U}{2L} \cre{c}_0 \cre{c}_0 \des{c}_{L/4} \des{c}_{L/4} + H.c. \right)
\label{heff}
\end{equation}
where $\op{n}_q = \cre{c}_q \des{c}_q$ are the number operators and the parameter $\eta$ evaluated at $\theta=\pi/4$ is given by $\eta = \frac{J^2-t^2}{J^2+t^2}$.
The first bracket of Eq.~(\ref{heff}) contains the contributions from the single-particle Hamiltonian (\ref{hamspdiag}), whereas the terms in the second and third brackets arise from the on-site interaction.
At the critical phase twist $\theta = \pi/4$ the former are an unimportant zero-energy offset, whereas the terms in the second bracket shift the energies of the states in the subspace differently, e.g.~they lead to an energy difference of $U(N-1)/L$ between the states $\ket{N,0}$ and $\ket{N-1,1}$, while the states $\ket{n,N-n}$ and $\ket{N-n,n}$ remain pairwise degenerate.
The terms in the third bracket are off-diagonal in the Fock basis of the subspace and describe two-particle scattering between the two single-particle modes.

Let us now determine the ground and first excited state for slightly non-uniform rings $t/J \approx 1$, close to the critical phase twist $\theta \approx \pi/4$. Since the terms in the second bracket of Eq.~(\ref{heff}) increase the energy for all states in the subspace apart from $\ket{N,0}$ and $\ket{0,N}$ and the coupling between the states is weak (as the coupling $\eta$ is small in this limit), we project the effective Hamiltonian (\ref{heff}) onto the subspace spanned by these two nearly-degenerate lowest-energy states.
As there is no direct coupling between $\ket{N,0}$ and $\ket{0,N}$ we calculate the total coupling through intermediate states using perturbation theory. After eliminating the intermediate states we obtain the following $2 \times 2$ Hamiltonian
\begin{equation}
\op{H}_{2 \times 2} = \left( \begin{array}{cc}
\Delta E/2 & \Delta \\
\Delta^* & - \Delta E/2
\end{array}
\right)
\label{ham2by2}
\end{equation}
where $\Delta E$ is the energy difference between the states $\ket{N,0}$ and $\ket{0,N}$ caused by the detuning of the phase twist from resonance $\Delta \theta = \theta - \pi/4$, i.e. $\Delta E = N(E_{L/4}^- - E_0^-) \approx \frac{4J t N \Delta \theta}{\sqrt{J^2 + t^2}}$, and $\Delta$ is the coupling between the states $\ket{N,0}$ and $\ket{0,N}$ due to the off-diagonal terms of the effective Hamiltonian (\ref{heff}). As the latter only directly couples the states $|n,N-n\rangle$ and $|n\pm2,N-n\mp 2\rangle$, the first non-vanishing order is given by
\begin{equation}
\Delta = \frac{\bra{N,0} \op{H}_{\textrm{eff}}^{N/2} \ket{0,N}} {\prod_{j=1}^{N/2-1}(E_0^{(1)}-E_{2j}^{(1)})} = \frac{U}{L} \cdot \left(\frac{i \eta}{2}\right)^{N/2} \cdot \frac{N!}{\prod_{j=1}^{N/2-1} (2j)^2}
\label{gap2}
\end{equation}
with the interaction energy shift $E_n^{(1)} = \frac{U}{2L} \left( 2 n (N-n) + N^2 - N \right)$.
Note that in contrast to the case of the ring lattice the perturbation couples states within the degenerate manifold (see Fig.~\ref{fig:coupling}). This leads to a less severe but still exponential scaling of the gap $\Delta$ with the number of particles $N$ \citep{Nunnenkamp08}.

The ground state of the two-by-two Hamiltonian (\ref{ham2by2}) is similar to the one obtained above in Eq.~(\ref{cat}), i.e.
\begin{equation}
\frac{\alpha \ket{N,0} + i^{N/2} \beta \ket{0,N}}{\sqrt{2}}.
\end{equation}
We see that to obtain a cat-like superposition, i.e.~$\alpha/\beta \approx 1$, the energy difference must not dominate over the coupling $|\Delta E| \ll |2\Delta|$ (see \cite{Hallwood06b} and \cite{Nunnenkamp08} for further details).

\section{Vortex nucleation in the lowest Landau level}

In this section, we discuss the effects of rotation on ultracold bosons confined to a two-dimensional harmonic potential. We first review the results of \cite{Dagnino2009a, Dagnino2009b} in order to then compare and contrast them with our findings on rotating ring (super)lattices which we presented in the previous section.

\subsection{Hamiltonian}

Following \cite{Dagnino2009a, Dagnino2009b}, we consider a system of $N$ ultracold bosons with mass $M$ confined to a two-dimensional symmetric harmonic potential $V_0$ and rotating in the $x-y$ plane about the $z$-axis with angular velocity $\Omega$, then as before the many-body Hamiltonian in the rotating frame is
\begin{equation}
\hat {H}=\int d\mathbf{x}
\hat{\Phi}^{\dagger}\left[-\frac{\hbar^2}{2 M}\nabla^2 +
V_0(\mathbf{x}) +\frac{4 \pi \hbar^2 a_s}{2M}
\hat{\Phi}^{\dagger}\hat{\Phi}- \Omega \hat{L}_z\right]\hat{\Phi}.
\end{equation}

If we assume that the potential $V_0$ confines the motion along the $z$-axis so strongly that only the transversal motion of the atoms in the $x-y$ plane needs to be considered, and that along the transverse direction the atoms feel a harmonic confinement $V_0(x,y) = M \omega^2 (x^2 + y^2)/2$ and the interaction energy is small compared to the Landau level splitting $\hbar (\omega + \Omega)$, the bosonic field operator $\hat{\Phi}$ can be expanded in the lowest Landau level basis $\hat{\Phi} = \sum_m \hat{a}_m \varphi_m (x,y)$. Here, $\varphi_m(x,y)$ are the eigenfunctions of the single-particle angular momentum operator $\op{L}_z$ with non-negative integer eigenvalue $m$, i.e.~$\varphi_m(x,y)\propto (x + i y)^m e^{-(x^2 + y^2)/2 \lambda^2}$ with the magnetic length $\lambda = \sqrt{\hbar/M\omega}$. In the course of the following discussion an asymmetry of the trapping potential in the $x-y$ plane will be included by adding a single-particle potential $V(x,y) = 2 A M \omega^2 (x^2-y^2)$, where $A$ is a measure of the asymmetry. For $A \ll 1$ the lowest Landau level remains a good basis set and the asymmetry can treated as a perturbation.
Within this approximation the many-body Hamiltonian is \citep{Dagnino2009a, Dagnino2009b}
\begin{align}
\op{H} & = \op{H}_0 + \op{U} + \op{V} \nonumber \\
& = \hbar (\omega - \Omega) \op{L}
+ \frac{g}{4\pi \lambda^2} \sum_{ijkl} \frac{(k+l)! \delta_{i+j,k+l}}{2^{k+l} \sqrt{i!j!k!l!}} \cre{a}_i \cre{a}_j \des{a}_k \des{a}_l \nonumber \\
& + \frac{A}{2} \lambda^2 \sum_m \sqrt{m(m-1)} \cre{a}_m \des{a}_{m-2} + \sqrt{(m+1)(m+2)}\cre{a}_m \des{a}_{m+2}.
\label{totalHam}
\end{align}
Here, $\op{H}_0$ is the unperturbed single-particle Hamiltonian proportional to the total angular momentum operator $\op{L} = \sum_m m \cre{a}_m \des{a}_m$, $\op{U}$ is the two-body interaction and $\op{V}$ is the perturbation due to the asymmetry in the single-particle potential. In the following we use $\lambda$, $\hbar \omega$ and $\omega$ as units of length, energy and frequency, respectively.

\subsection{Effective many-body Hamiltonian}

\begin{figure}
\centering
\includegraphics[width=\textwidth]{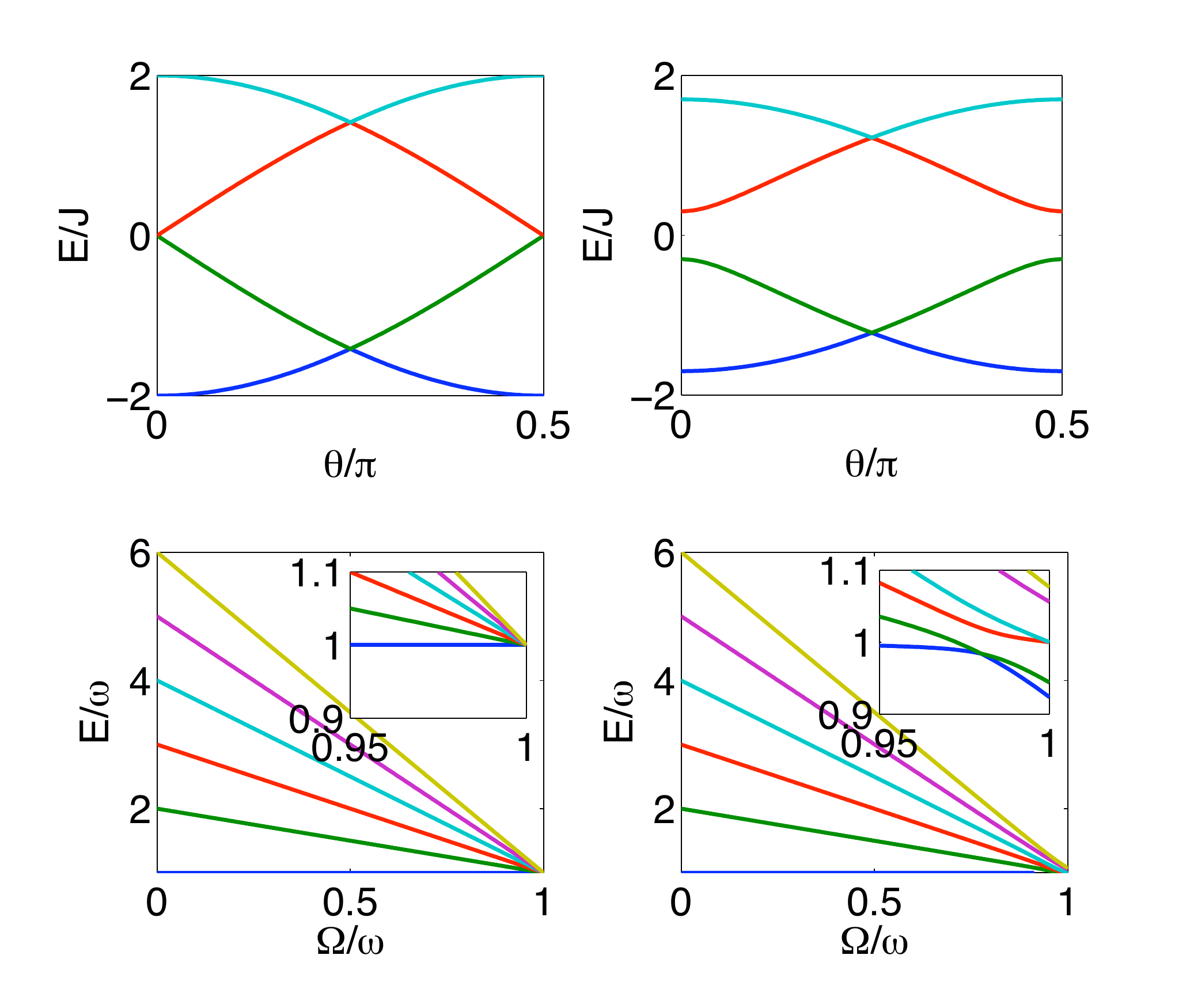}
\caption{{\bfseries Single-particle spectrum versus phase twist (or rotation frequency).}
(Upper) Rotating ring lattice with $L=4$ sites: there is a single-particle level crossing at $\theta = \pi/4$ both for $t=J$ (left) and $t=0.7J$ (right) \citep{Nunnenkamp08}.
(Lower) LLL system: both for $A = 0$ (left) and $A = 0.03$ (right) the single-particle spectrum is degenerate only at $\Omega = \omega$ for $A = 0$ (left) or at $\Omega \approx \omega$ for $A = 0.03$ (right). This crossing at single-particle level is not important for the discussion that follows. The insets show the details around $\Omega \approx \omega$.}
\label{fig:spspectrum}
\end{figure}

Analogous to the previous section, we start our discussion with the single-particle spectrum $E_m = \hbar(\omega-\Omega) m$. In Fig.~\ref{fig:spspectrum} we plot it as a function of rotation frequency $\Omega$. We see that all states in the lowest Landau level become degenerate at $\Omega = \omega$, but in contrast to the rotating ring lattice there is \emph{no single-particle level crossing} for $0 \le \Omega < \omega$.

Nonetheless, the spectrum of $N$ non-interacting bosons is highly degenerate for any rotation frequency $\Omega$. This is a consequence of the fact that the single-particle Hamiltonian is not only diagonal in the angular momentum basis but proportional to the total angular momentum $L$, so that all many-body states with the same total angular momentum $L$ have the same energy -- a degeneracy independent of the rotation rate $\Omega$ and only present at the many-body level. Interactions will break this huge degeneracy, but since the interaction Hamiltonian $\op{U}$ commutes with the total angular momentum operator $\op{L}$ the interacting many-body Hamiltonian remains block diagonal, i.e.~it only mixes states with the same total angular momentum $L$.

\cite{Smith2000} have shown that the states $\ket{\Phi_L}$ with the wave functions
\begin{equation}
\Phi_L(x_1,y_1,\dots,x_N,y_N)
\propto \sum_{1 \le i_1 \le \dots i_N \le N} (z_{i_1} - z_0) \dots (z_{i_L} - z_0) \, \Phi_0 (x_1,y_1,\dots,x_N,y_N)
\end{equation}
where $z_j = x_j + i y_j$ are complex coordinates in the $x-y$ plane, $z_0 = \sum_j z_j/N$ is the center of mass, and $\Phi_0 \propto e^{- \sum_j z_j^2/2}$, i.e.~the non-rotating ground-state wave function, are exact eigenstates of the interaction Hamiltonian $\op{U}$ with eigenvalue
\begin{equation}
E_L = \frac{g N}{8 \pi} \left( 2N - L - 2\right).
\end{equation}
They also present overwhelming numerical evidence that these are the ground states for $0\le L \le N$ with $L \not= 1$.
When expressed in the Fock basis $\ket{n_0,n_1,\dots}$ with occupation numbers $n_m$ for the angular-momentum basis, these are seen to be complicated superposition states. For example, the many-body state with one unit of angular momentum per particle $\ket{\Phi_N}$ is not, in fact, the vortex state $\ket{0,N,0,\dots}$ with all particles occupying the single-particle wave function with $m=1$, but rather the so-called yrast state whose occupation number of the first angular-momentum state is large $\langle n_1 \rangle \approx N$ but whose overlap with the vortex state is only about one half \citep{Bertsch1999}.

\begin{figure}
\centering
\includegraphics[width=\textwidth]{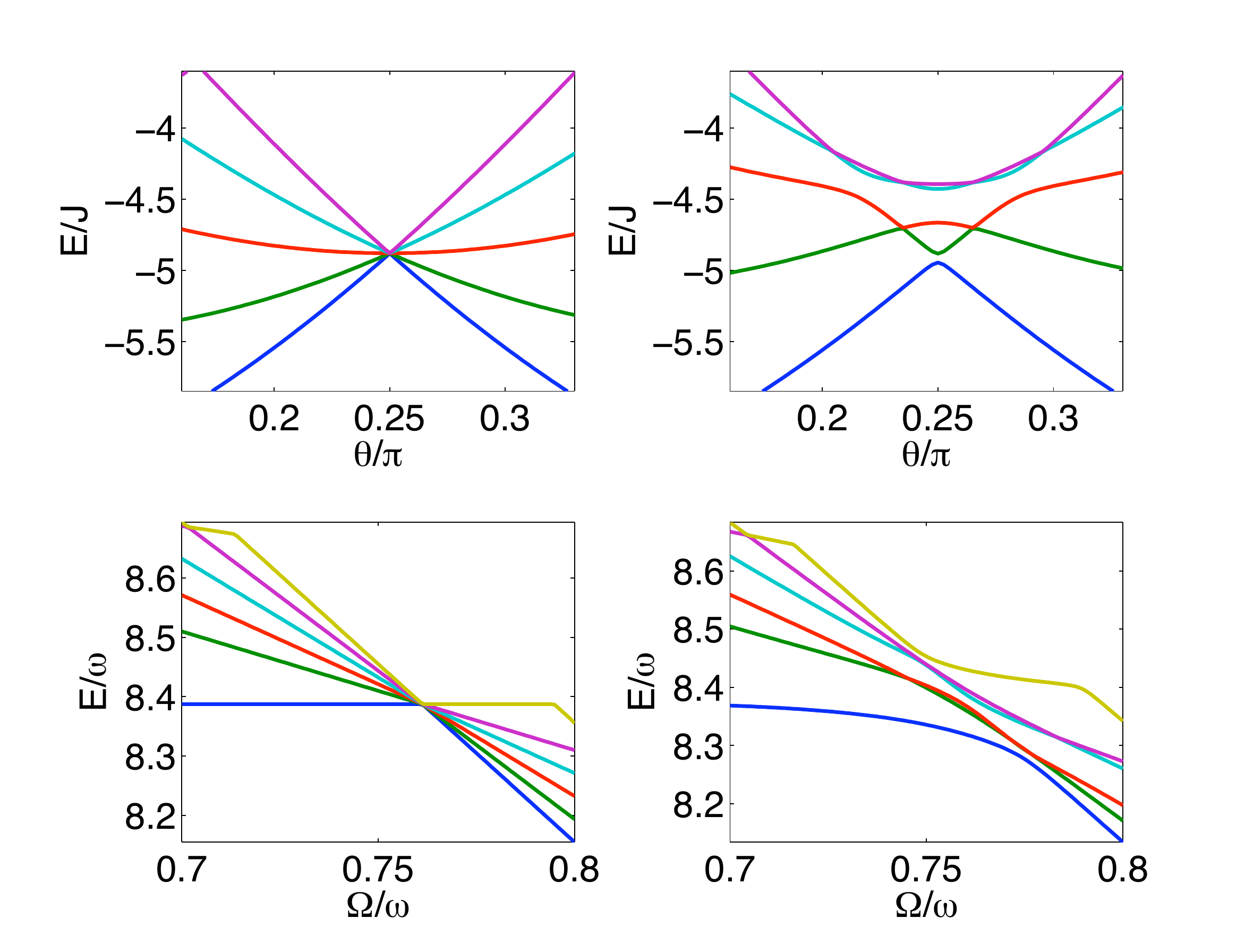}
\caption{{\bfseries Many-particle spectrum versus phase twist (or rotation frequency).} (Upper) In the rotating ring lattice with $L = N = 4$ and $t=0.7J$ the $(N+1)$-fold degeneracy of the non-interacting system (left) is lifted to an anti-crossing in the interacting system $U/J=0.5$ (right) \citep{Hallwood06a}. (Lower) In the LLL system with $N=6$ particles the $N$-fold degeneracy of the interacting isotropic ($g = 1$ and $A = 0$) case (left) is lifted in the asymmetric ($g = 1$ and $A = 0.03$) case (right) \citep{Dagnino2009a}.}
\label{fig:spectrum}
\end{figure}

Following \cite{Dagnino2009a}, we plot in Fig.~\ref{fig:spectrum} the many-body spectrum as a function of rotation frequency $\Omega$. Since the single-particle and interaction energy of the states $\ket{\Phi_L}$ depend linearly on $L$, there is a rotation frequency $\Omega_c / \omega = 1 - \frac{g N}{8\pi}$ at which all many-body states $\ket{\Phi_L}$ are degenerate. Around this rotation frequency the ground state rapidly changes: for $\Omega < \Omega_c$ the ground state is $\ket{\Phi_0}$, while above $\Omega > \Omega_c$ it is the yrast state $\ket{\Phi_N}$.

Turning to the effect of the perturbation $\op{V}$ on the states involved, we plot in Fig.~\ref{fig:spspectrum} the single-particle spectrum as a function of rotation frequency $\Omega$. We see that apart from a single-particle crossing close to $\Omega \approx \omega$ the perturbation has little effect on the single-particle spectrum, and it turns out that this level crossing is unimportant for the discussion that follows.

The many-body spectrum, on the other hand, is degenerate at $\Omega_c$, so that any additional perturbation can effectively mix the degenerate many-body states $\ket{\Phi_L}$. Like \cite{Dagnino2009a}, we show in Fig.~\ref{fig:spectrum} the many-body spectrum for $A \not= 0$, where we find the degeneracy is lifted into a set of anti-crossings.

At first glance the situation seems to be similar to the case of the superlattice discussed in the previous section. In both cases the $N+1$-fold ($N$-fold) degeneracy in a set of many-body states is lifted by an external perturbation (interactions in the superlattice case and trap asymmetry in the LLL case) which couples the various many-body states among them. This is not the case in the uniform ring lattice where coupling takes place via non-resonant states (see Fig.~\ref{fig:coupling}).

\begin{figure}
\centering
\includegraphics[width=\textwidth]{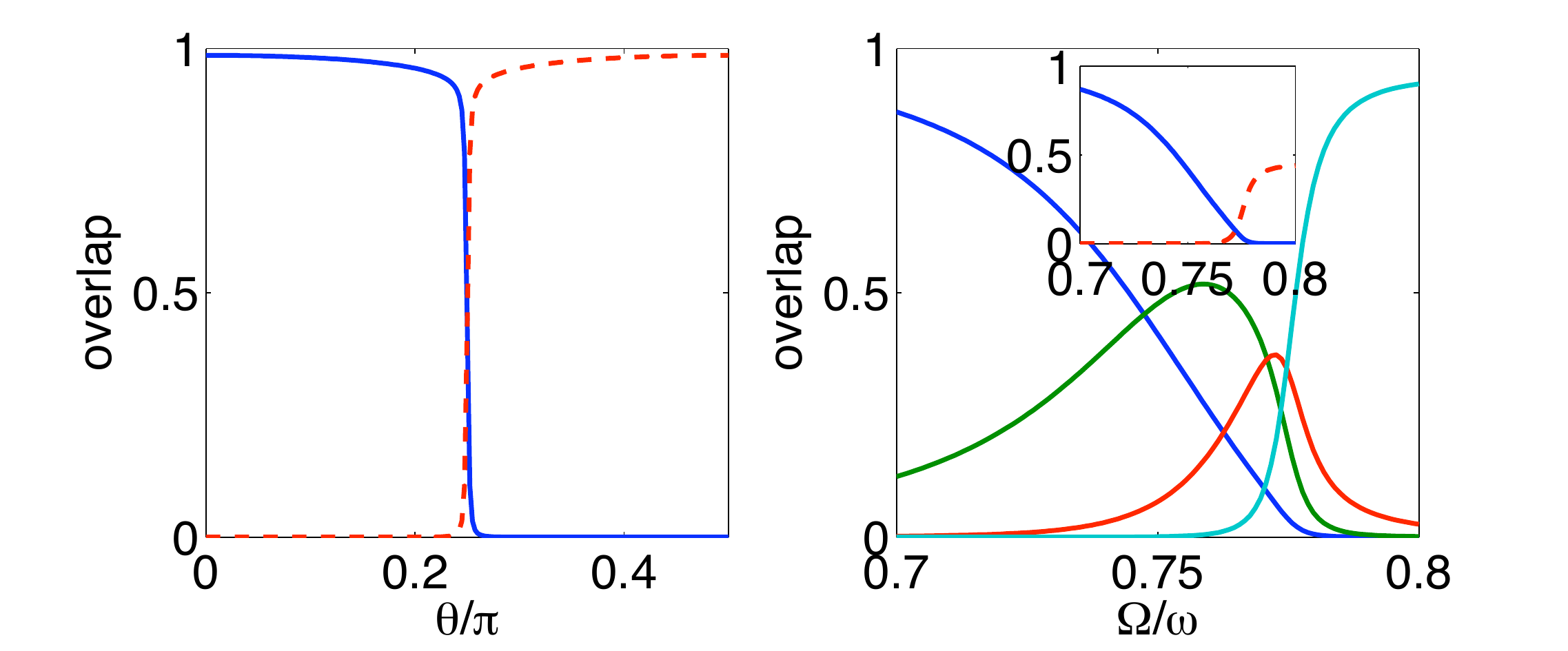}
\caption{{\bfseries Properties of the ground-state wave function.} (Left) Overlap of the ring lattice ground state with the wave function of a condensate in the two lowest-lying quasi-momentum states: away from the critical phase twist the atoms condense \citep{Rey07}. (Right) Overlap of the LLL ground state for $N=6$ particles in the asymmetric case ($g = 1$ and $A = 0.03$) with the basis set $|\Phi_L\rangle$ of the effective Hamiltonian: it changes from $|\Phi_{0} \rangle$ via the superposition (\ref{maxentstate}) to the yrast state $|\Phi_{N} \rangle$ \citep{Dagnino2009a}. We use exact diagonalization of the many-body Hamiltonian (\ref{totalHam}) including all states up to a total angular momentum of $L_\mathrm{max} = N + 2$. For comparison the inset shows the overlap of the LLL ground state with non-rotating condensate (solid) and vortex (dashed) states.}
\label{fig:overlap}
\end{figure}

However, there are two crucial differences between the two systems. First of all, \emph{the interaction Hamiltonian $\op{U}$ does not simply couple the degenerate states but also shifts their energies differentially} (see Fig.~\ref{fig:coupling}). This is why we were able to construct an effective two-by-two Hamiltonian $\op{H}_{2 \times 2}$ (\ref{ham2by2}) that only couples the two lowest-lying states, and found the ground-state to be a macroscopic superposition state exhibiting an energy gap which exponentially decreases with increasing number of particles (due to virtual couplings to the adiabatically eliminated states within the resonant manifold). In contrast, \emph{the perturbation $\op{V}$ contains only off-diagonal matrix elements} and all states within the degenerate manifold have to be treated on an equal footing. This view is corroborated in Fig.~\ref{fig:overlap} where we, like \cite{Dagnino2009b}, plot the overlap of the ground state with the states $\ket{\Phi_L}$ as a function of rotation frequency $\Omega$. We see that with increasing $\Omega$ the weight of the ground state in the various $\ket{\Phi_L}$ states shifts towards higher $L$ in steps of two units of angular momentum.
At the nucleation point \cite{Dagnino2009a} find that the state of the system is well described by a complicated superposition state
\begin{equation}
\ket{\Psi_0} \propto \ket{N,0} + \ket{N-2,2} + \dots + \ket{2,N-2} + \ket{0,N}
\label{maxentstate}
\end{equation}
where $\ket{n,m}$ is the state with $n$ and $m$ atoms in the two eigenfunctions of the single-particle density matrix $n^{(1)}(r,r') = \langle \Psi_0 | \cre{\Psi}(r) \des{\Psi}(r') | \Psi_0 \rangle$ with the largest eigenvalues, respectively.
The nature of the intermediate states also leads to a different scaling of the energy gap with the number of particles. As is shown by \cite{Dagnino2009a}, the energy gap within the subspace at the critical rotation frequency remains finite as the number of particles increases, as opposed to the exponential scaling of the energy gap for the macroscopic superposition states in the ring lattice. Note, however, that the energy difference between the ground state within the subspaces of even and odd $L$ decreases exponentially with the number of particles \citep{Parke2008}.

The second difference between vortex nucleation in the two systems is the ground state above the critical rotation frequency. In the LLL system it has a large overlap with the yrast state $\ket{\Phi_N}$. To compare with the superlattice system we plot in Fig.~\ref{fig:overlap} also the overlap between the ground state and the macroscopic occupied modes on both sides of the resonance: i.e.~the condensate at rest $|N, 0, \dots \rangle$ for $\Omega<\Omega_c$ and the vortex state $|0, N, \dots \rangle$ for $\Omega>\Omega_c$. We see that whereas the overlap with the corresponding states is always close to one for the superlattice, in the LLL the overlap is only close to one for $\Omega<\Omega_c$. For $\Omega>\Omega_c$ is only about one half. This is a consequence of the fact that in the ring lattice away from the critical rotation frequency the state $\ket{0,N}$ is the only low-lying state and therefore finite interactions lead to a small depletion of the superfluid ground state. In contrast, the non-interacting LLL system with one unit per particle is highly degenerate and interactions produce strongly-correlated many-body ground-states like $\ket{\Phi_N}$.

\section{Noise Correlations}

To further quantify the entanglement of the ground state at the critical frequency and to compare the different nature of the ground state in the ring lattice and LLL system we calculate the noise correlation pattern, i.e.,
\begin{equation}
\Delta (q,q') = \langle \hat{O}^\dagger_{q'} \hat{O}_{q'} \hat{O}^\dagger_q \hat{O}_q\rangle - \langle \hat{O}^\dagger_{q'} \hat{O}_{q'} \rangle  \langle \hat{O}^\dagger_q \hat{O}_q \rangle.
\end{equation}
We will use quasi-momentum creation operators for the lattice case $\hat{O}_{q} =\hat{b}_{q}$ and angular-momentum operators for the harmonically trapped system, $\hat{O}_{q} = \hat{a}_{q}$.

\cite{Mintert2009} have argued that noise correlations are experimentally accessible quantities encoding information on entanglement. An alternative characterization of the entanglement properties of the ground state in the LLL system has been given by \cite{Liu2009}, and \cite{Read2003} have proposed time-of-flight expansion to probe the vortex lattice in LLL systems.

For the lattice and superlattice cases the noise interferogram shows the development of three sharp fringes, with positive and negative signs at the critical frequency. Their amplitudes are given by
\begin{equation}
\Delta (0,0) = \Delta (1,1) = -\Delta (0,1) = N^2/2
\end{equation}
heralding the development of a Schr\"odinger cat state at $\Omega_c$. Away from $\Omega_c$ the noise interference pattern disappears signaling the unentangled nature of the ground state and the macroscopic occupation of a single mode \citep{Rey07}.

\begin{figure}
\centering
\includegraphics[width=0.9\textwidth]{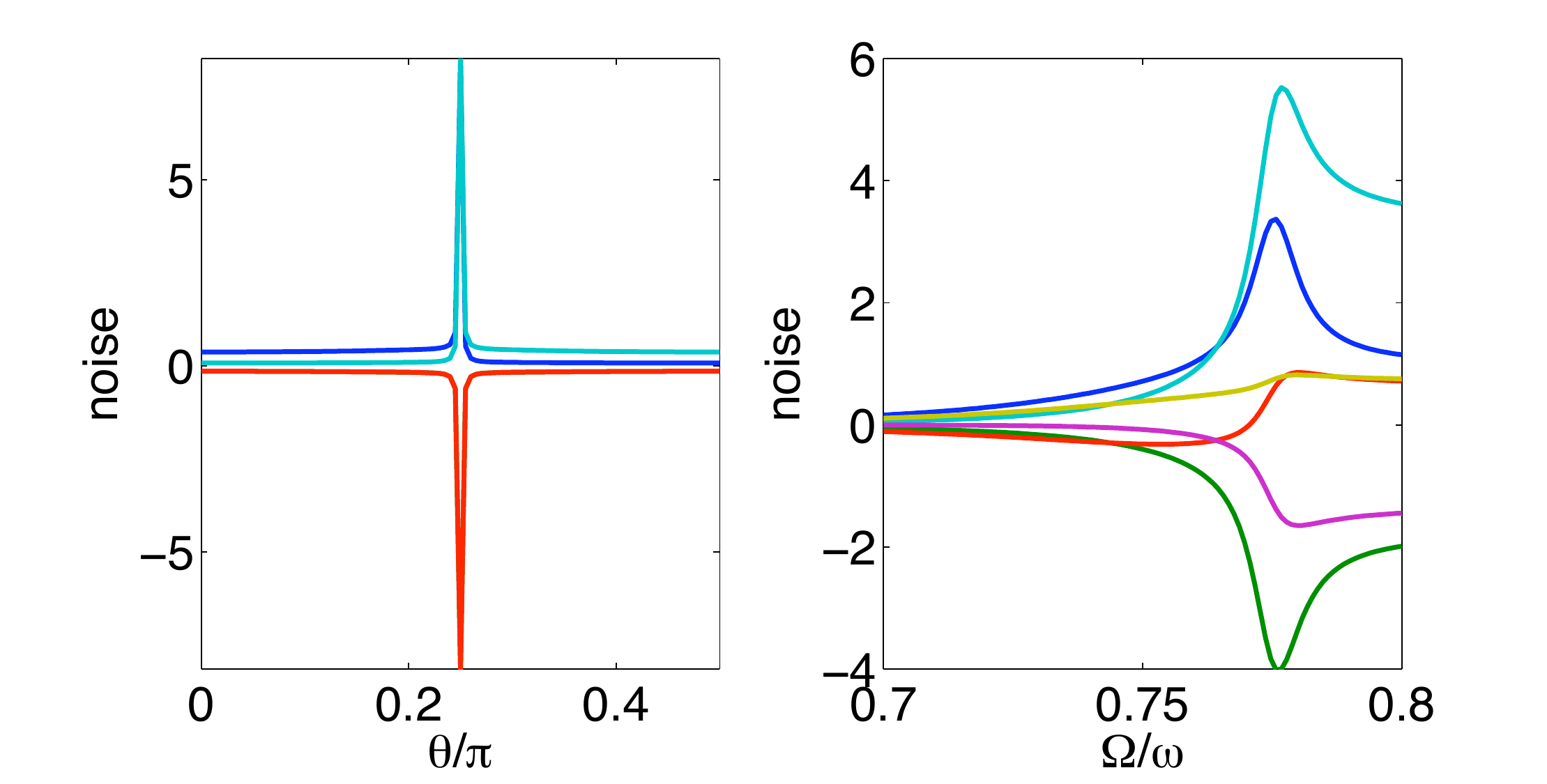}
\caption{{\bfseries Quasi-momentum (or angular-momentum) noise correlations.} The two different strongly-correlated ground states can be distinguished via noise correlations: in the first case (left) \citep{Rey07} they are non-zero only on resonance, while in the latter (right) they are strongest on resonance but non-zero above resonance as well. This corroborates the fact that the atoms do not condense into one single-particle state but form a strongly-correlated many-body state.}
\label{fig:noise}
\end{figure}

In striking contrast is the noise interferogram for the LLL system which exhibits various sharp fringes which develop as the system is driven through the critical frequency. They signal the strong correlations and multi-mode nature of the ground state at $\Omega_c$. Furthermore, the interferogram does not disappear for $\Omega>\Omega_c$, behavior which highlights the correlated nature of the yrast state $\ket{\Phi_N}$. All these features are illustrated in Fig.~\ref{fig:noise} where we plot the non-zero $\Delta (q,q')$ as a function of phase twist $\theta$ and rotation frequency $\Omega$, respectively.

\section{Conclusion}

We have shown how a new generation of experiments can be used to examine the quantum nucleation process and its important link to entangled states of atoms. We have compared quantum vortex nucleation in rotating ring lattices and two-dimenisonal harmonic potentials and have shown how effective Hamiltonians can be used to describe the nearly-degenerate ground-state manifold close to the rotation frequency at which the first vortex is nucleated. The degeneracy in the many-body spectrum is lifted by interactions in the case of the ring superlattice and by the asymmetry of the single-particle potential in the LLL system.
In the first case the interactions not only couple the states but also shift their energies differentially. This leads to macroscopic superpositions as low-lying states whose energy gap decreases exponentially with increasing number of particles. In contrast, in latter case all states contribute to the ground-state wave function and the energy gap within the subspace remains finite with increasing particle number. Finally, we showed that the two scenarios can be distinguished in noise correlation interferograms: while the ground state in a rotating ring lattice above the critical phase twist is the vortex state with vanishing noise correlations, interactions in the lowest Landau level produce the yrast state which has a non-trivial noise pattern.

\section*{Acknowedgements}

AN thanks Steven M.~Girvin for insightful discussions and a careful reading of the manuscript. AMR acknowledges support from NSF, NIST and the DARPA OLE program and AN acknowledges support from NSF DMR-0603369.

\label{lastpage}

\end{document}